\documentclass{article}
    \usepackage[margin=3cm]{geometry}
    \usepackage{amsmath}

\title{A new classical solution in SU(2) Yang-Mills gauge theory}
\author{H. Albert\\Physics Department\\University of Hamburg\\ \footnotesize{Hinnerk.Albert@gmx.de} }

\begin{document}

    \maketitle
    \begin{abstract}
  	  A new solution of euclidean Yang-Mills gauge theory, which is governed by $\pi_4 (SU(2)) $, is given . Its relationship to knot theory and Hopfions is discussed
    \end{abstract}

\section{Introduction}
A new classical solution of the SU(2) Yang-Mills gauge theory is given. It is based on the homotopy group
\begin{equation}
\fbox{$\pi_4 (SU(2))= \pi_4 (S^3 ) = Z_2 $}
 \end{equation}
  Various ansaetze are discussed, either selfdual as in the case of instantons~\cite{Be}with finite action but zero energy or in Minkowski space with finite energy  as in the meron case~\cite{Fu} We will also investigate whether the Corrigan-Fairlie ansatz works, i.e., finding a $\phi^4 $ solution and plugging it into the general ansatz given by them~\cite{Cr}

\section{Main part}
In four dimensional euclidean Yang-Mills gauge theory, the base space (refering to the language of principal fiber bundles) $R^4 $ can be compactified to the fourdimensional sphere $S^4 $ because pure Yang-Mills gauge theory is conformally invariant and the metric of the fourdimensional sphere is conformally flat:
\begin{equation}
g_{\mu \nu } = \Omega (x) \eta_{\mu \nu }
\end{equation}
 where g is the metric of the sphere $S^4$, $\eta$ the metric of $R^4 $ (flat metric ) and $\Omega $ the conformal transformation ~\cite{Si}. Since the local gauge transformations g (or gauge transformations of the second kind) define mappings from the base space of the principal fiber bundle into its structure group SU(N) (say), they are classified by $\pi_4 (SU(N))$ which is equal to $Z_2 $ in the case N = 2, otherwise 0 ~\cite{Pu}. In the case of instantons ~\cite{Ra} and merons one employs  boundary conditions, that enforce mappings from the equator of the base space $S^4 $, that is $S^3 $ , to SU(N). These are classified by $\pi_3 (SU(N)) = \pi_3 (SU(2) = \pi_3 (S^3 ) = Z$ because of Bott periodicity (for instance D.Husemoller, fiber bundles ) The solutions are then classified by elements of $\pi_3 (S^3 ) = Z$, which is called their topological charge k. In the case of instantons, selfduality of the curvature tensor $F_{\mu \nu } $ is demanded, to have finite action solutions (and zero energy), because it is believed, that these dominate the euclidean path integral in the case of semiclassical calculations ~\cite{Co}. The instanton solution ~\cite{Bel} with k = 1 is given by
 \begin{equation}
 A_{\mu} = \frac{x^2 }{x^2 + \lambda^2 } g^{-1} \partial_{\mu} g
  \end{equation}
  where $\lambda$ is the scaling factor (pure Yang-Mills gauge theory is scale invariant and for merons with k = 1
  \begin{equation}
  A_{\mu} = \frac{1}{2} g^{-1 } \partial_{\mu} g
   \end{equation}
   (due to the factor 1/2 this is not a pure gauge, hence $F_{\mu \nu } $ is not zero, and gives a solution with finite energy~\cite{Fu}.) (We can generalize the factor $\frac{1}{2}$ to $\frac{m}{n}$ with $m \neq n$ ) The gauge transformations g are representants of $\pi_3 (S^3 ) $, and hence mappings from$S^3$ to $S^3$ with mapping degree 1 while $g^{-1} \partial_{\mu} g $ is the pullback (see Bott, Differential forms in Algebraic Topology) Explicit formulas for the representants shall be given in a forthcoming paper.
To find first solutions one can replace the [g] of $\pi_3 (S^3 ) $ with [g] out of $\pi_4 (S^3 ) $ in the instanton and meron solution ([g] means equivalence class of the mapping g) . These have then topological charge out of $Z_2 $. Further solutions can be found by employing the theorem of Corrigan and Fairlie ~\cite{Cr}: $A_{\mu} = i\sigma_{\mu \nu } \partial_{\nu} ln \phi $ is a solution of the Yang-Mills gauge theory, if $\phi $ is a solution of the $\phi^4 $ theory. There is a relationship between $\pi_4 (S^3 )$ and braid groups ~\cite{Wu}. Also there is a lot to be said about $\pi_4 (SU(2))$ itself ~\cite{Pu}. These two items and further solutions shall be investigated in a forthcoming paper. An explicit, nontrivial representant of the homotopy group $\pi_4 (SU(2)) $ can be found by suspending the Hopf map, i.e., a representant of $\pi_3 (SU(2)) $ ~\cite{Fr}.
Although $\pi_4 (SU(N)) $ is zero for N larger than 2 ~\cite{St}, the author wonders whether there is any connection to the confinement problem and triality ($Z_3 $ vortices).
An explicit representation of the nontrivial element of $\pi_4 (S^3 )$ was given by D.Friedan ~\cite{Fr} based on a paper written by T.Puettmann and Rigas ~\cite{Pu}. In what follows, we will describe this mapping and find an interesting reformulation of it, linking it to Skyrmes Hedgehog solution ~\cite{Sk}. For the ansatz of the vector potential $A^a_{\mu}$ we had chosen the generalized meron ansatz
\begin{equation}
A_{\mu} =\frac{m}{n} \partial_{\mu} g g^{-1}
 \end{equation}
  The pre factor $\frac{m}{n}$ guarantees a non vanishing field strength $F_{\mu \nu}$. For convenience and without breaking generality we choose this pre factor as $\frac{1}{\sqrt{8}}$. The Yang-Mills action then reads as
\begin{equation}
 S = \frac{1}{32} \int dx^4 [\partial_ {\mu } g g^{-1} , \partial_{\nu} g g^{-1} ]^2
\end{equation}
 which is exactly the term Skyrme added to the nonlinear $\sigma$ model to stabilize soliton solutions. The mapping g will be given now
\begin{equation}
g (x,z) = \cos (\varphi (x) )
\left(
\begin{array}{*{4}{c}}
1 & 0 \\
0 & 1 \\
\end{array}
\right)
+ i\sin (\varphi (x) )
\left(
\begin{array}{*{4}{c}}
|z_1|^2 - |z_2 |^2 & 2 z_1 \bar{z_2 } \\
2\bar{z_1 } z_2 & -|z_1 |^2 + |z_2 |^2 \\
\end{array}
\right)
\end{equation}
where z = ($z_1 , z_2 ) \epsilon C^2$ and x $\epsilon R $ and $z_1 = x_1+ix_2$ and $z_2 = x_3 + i x_4$. Evaluating the mapping g(x,z) as an exponent according to the Lie algebra of SU(2), i.e., the Pauli matrices $\sigma^a $, a =1,2,3, g(x,z) can be written as follows:
\begin{equation}
g(x,z ) = \exp[i\sigma^a \zeta^a \varphi (x) ]
\end{equation}
 where the $\zeta^a$ constitute the Hopf map
 \begin{equation}
 f:S^3 \rightarrow S^2
 \end{equation}
  with Hopf invariant 1 (Heinz Hopf, Collected papers, Springer, Berlin) with
\begin{equation}
\zeta_1 = 2(x_1 x_3 + x_2 x_4 ) = \sin \theta \cos \varphi
\end{equation}
 \begin{equation}
 \zeta_2 = 2(x_1 x_4- x_2 x_3 ) = \sin \theta \sin \varphi
 \end{equation}
 \begin{equation}
  \zeta_3 = x_1^2 + x_2^2 - x_3^2 - x_4^2 = \cos \varphi
  \end{equation}
  The $x_i$ , i = 1,2,3,4, are the coordinates of the Euclidean space, in which $S^3$ is embedded.
$\varphi (x)$ obeys the following boundary conditions:
\begin{equation}
\varphi (-\infty ) = \pi
\end{equation}
 \begin{equation}
 \varphi (\infty ) =0
 \end{equation}
  ($\varphi $ is often called the chiral angle.) This representation reflects better the nonabelian character of the mapping g(x,z).
We then have
\begin{equation}
g(x, z) = \cos (\varphi (x) )+i(\sigma^1 \zeta_1 (z)+ \sigma^2 \zeta_2 (z) + \sigma^3 \zeta_3 (z) ) \sin (\varphi (x))
\end{equation}
 This is the formula, we will use to determine the gauge potential and the field strength and it is Skyrmes Hedgehog field ~\cite{Sk}. The function $\varphi(x)$ and the angle $\varphi$ in formulas 9, 10 and 11 are different objects. It is Skyrmes hedgehog solution on a second level, because the $\zeta^a $ coordinates have to be replaced by the $x^i $, i = 1, 2, 3, 4, coordinates. Hence, in contrast to the Skyrme hedgehog solution, our new solution is explicitly (Euclidean) time dependent.
The solution for the vector potential is then
\begin{equation}
A_i =\frac{1}{2}(i\sigma_i \sin (\varphi ) \cos (\varphi ) + \zeta_i +i \epsilon_{jik} \zeta_j \sigma_k (\sin (\varphi ))^2 )
\end{equation}
for i = 1,2,3 and
\begin{equation}
A_4 = i\frac{1}{2} \varphi ' (x) \sigma_a \zeta_a
\end{equation}
 with and $x \epsilon R$. A comment is in order concerning the prefactor $\frac{1}{2}$ . We had chosen this prefactor to be as close as possible to the meron and Skyrme model. But the ansatz will give a solution for $\frac{n}{m}$, n,m $\epsilon$ R, R the real numbers.
\\
$\varphi (x)$ can be determined by solving the equation of motion
\begin{equation}
\partial_{\mu} [A,A] + [[A,A]A] = 0
\end{equation}
 or by minimizing the Lagrangian with respect to $\varphi (x)$ :
 \begin{equation}
 \delta L (\varphi (x)) = 0
 \end{equation}

Minimizing L with respect to $\varphi (x)$ gives a Bernoulli differential equation.
\begin{equation}
p' (\varphi ) + \cot (\varphi ) p (\varphi ) + g (\varphi ) p^{-1} = 0
\end{equation}
with $p (\varphi ) = \varphi ' (x)$ and g($\varphi $) is a polynom function with respect to $\varphi$.
The solution of it is the profile function $\varphi (x)$:
\begin{equation}
\int \frac{d\varphi}{(A\sin^6 (\varphi ) + B\sin^4 (\varphi) + C\sin^2 (\varphi ) )^{1/2} } = x
\end{equation}
where A,B and C are polynoms purely in the coordinates $\zeta_i$. The integral, determining $\varphi $, is elliptic and must be solved approximately. Let F($\varphi )$ be the antiderivative of the elliptic integral above. Since F( $\varphi $) = x, where x is the suspension parameter, we have $F' (\varphi (x))$ = 1. Hence F is invertible and $\varphi $ can be determined as a function of x in principle.
This shows that the existence of the new solution of the SU(2) Yang-Mills gauge theory is ensured and completely determined. For a first approximation one may argue that the $\sin^2 (\varphi)$ term in the elliptic integral is dominant. Then the integral reduces to -$\int \frac{1}{\sin (\varphi) } = x$ (the minus sign coming from choosing the negative branch of the root in the denominator of the elliptic integral to fullfill the boundary conditions of $\varphi$ )This gives for $\varphi (x)$
\begin{equation}
\varphi (x) = 2\arctan (e^{-x} )
\end{equation}
The expression for $\varphi$ is called Lobachefskijs angle of Parallelism and plays a prominent role in hyperbolic geometry ~\cite{Gr} Plugging this into the ansatz for the vector potential gives
\begin{equation}
A_i = \frac{1}{2(\cosh (x))^2}(i\sigma_i \sinh (x) + \zeta_i + i\epsilon_{jik} \zeta_j \sigma_k )
\end{equation}
for i  1,2,3 and
\begin{equation}
A_4 = -i\frac{1}{\cosh(x)}\sigma_a \zeta_a
\end{equation}
This gives the potentials on $S^2 \times R$ . To gain the potentials on $S^4$, or better $S^3 \times R$, we have to plug in the expressions of the $\zeta$'s in terms of the $x_i$, i = 1,2,3,4 (equations 10,11,12). The reason, why we have a solution on $S^4$ minus the south- and north pole is, that the mapping $S^ \rightarrow S^3$ is gained from the Hopf mapping by suspension. But suspending $S^3$ gives not directly $S^4$ but a double cone with singularities on the north and south pole and $S^3$ as the equator. Cutting away the poles opens the cone to $S^3 \times R$, R being the real numbers. The final expression of the potentials $A_{\mu}$ are derived by stereographic projection.

An alternative method is to calculate $\varphi$ is using a socalled stick function as Skyrme used: Choose a so called stick function $\varphi (x) = \pi (1 - \lambda e^x )$  for x $< \lambda$ and zero else to fullfill the boundary conditions of $\varphi$. Plug this into the Lagrangian and variate with respect to $\lambda$. This gives in our case for $\lambda$ approximately
 \begin{equation}
 \lambda = 0.3533
 \end{equation}

 The potential $A_i$, i = 1,2,3,4, take then the following form
 \begin{equation}
 A_i = \sin^2 (\lambda e^x) [-i\cot (\lambda e^x ) + \zeta^a \sigma^a ]\sigma_i , i = 1,2,3
 \end{equation}
 \begin{equation}
 A_4 = - i \pi \lambda e^x \zeta^a \sigma^a
 \end{equation}
 From these we find the field strength tensors $F_{\mu \nu}$ for the case that $\varphi (x) = 2\arctan (e^{-x} )$:
 \begin{eqnarray}
 F_{12} = \frac{i}{2} \frac{1}{(\sinh (x))^4 } (-((\sinh (x))^2 + \zeta_3 \zeta_3 )\sigma_3 + (\zeta_2 \zeta_3 - \zeta_1 \sinh (x) )\sigma_2 - (\zeta_1 \zeta_3 + \zeta_2 \sinh (x)\sigma_1 )\\
 F_{13} = \frac{i}{2} \frac{1}{(\sinh (x))^4 } (-((\sinh (x))^2 + \zeta_2 \zeta_2 )\sigma_2 + (\zeta_2 \zeta_3 + \zeta_1 \sinh (x) )\sigma_3 - (\zeta_1 \zeta_2 - \zeta_3 \sinh (x) )\sigma_1 )\\
 F_{23} = \frac{i}{2} \frac{1}{(\sinh (x))^4 } (-((\sinh (x))^2 + \zeta_1 \zeta_1 )\sigma_1 + (\zeta_1 \zeta_2 + \zeta_3 \sinh (x) )\sigma_2 - (\zeta_1 \zeta_3 - \zeta_2 \sinh (x) )\sigma_3 )\\
 F_{14} = \frac{i}{2} \frac{1}{(\sinh (x))^2 }\frac{1}{\cosh (x)} (((-\zeta_2 \sinh (x)) + \zeta_1 \zeta_3 )\sigma_3 - (\zeta_1 \zeta_2 + \zeta_3 \sinh (x) )\sigma_2 - (\zeta_2 \zeta_2 + \zeta_3 \zeta_3 )\sigma_1 )\\
 F_{24} = \frac{i}{2} \frac{1}{(\sinh (x))^2 }\frac{1}{\cosh (x)} (((\zeta_1 \sinh (x)) + \zeta_2 \zeta_3 )\sigma_3 - (\zeta_1 \zeta_2 + \zeta_3 \sinh (x) )\sigma_1 + (\zeta_1 \zeta_1 + \zeta_3 \zeta_3 )\sigma_2 )\\
 F_{34} = \frac{i}{2} \frac{1}{(\sinh (x))^2 }\frac{1}{\cosh (x)} (((\zeta_1 \sinh (x)) - \zeta_2 \zeta_3 )\sigma_2 + (\zeta_1 \zeta_3 + \zeta_2 \sinh (x) )\sigma_1 + (\zeta_1 \zeta_1 + \zeta_2 \zeta_2 )\sigma_3 )
 \end{eqnarray}
 The solutions on $S^4$ (i.e. $S^3 \times R$) are gained plugging in the expressions of the $\zeta_i$ in terms of the $x_{\mu} , \mu =1,2,3,4$, see equations 10, 11, 12 above. For the stick function $\varphi (x) = \pi (1 - \lambda e^x ) $ we find for the field strength tensor $F_{\mu \nu} $ the following formulae:
 \begin{eqnarray}
 F_{12} = \frac{i}{2} (\sin (\pi \lambda e^x ))^4 (-(\tan (\pi \lambda e^x)^{-2} + \zeta_3 \zeta_3 )\sigma_3 + (\zeta_2 \zeta_3 - \zeta_1 (\tan (\pi \lambda e^x ) )^{-1} )\sigma_2 - (\zeta_1 \zeta_3 + \zeta_2 (\tan (\pi \lambda e^x ))^{-1} )\sigma_1 )\\
 F_{13} = \frac{i}{2}(\sin (\pi \lambda e^x ))^4  (-(\tan (\pi \lambda e^x)^{-2} + \zeta_2 \zeta_2 )\sigma_2 + (\zeta_2 \zeta_3 + \zeta_1 (\tan (\pi \lambda e^x ))^{-1} )\sigma_3 - (\zeta_1 \zeta_2 - \zeta_3 (\tan (\pi \lambda e^x ))^{-1}  )\sigma_1 )\\
 F_{23} = \frac{i}{2} (\sin (\pi \lambda e^x ))^4 ((-(\tan (\pi \lambda e^x))^{-2} + \zeta_1 \zeta_1 )\sigma_1 + (\zeta_1 \zeta_2 + \zeta_3 (\tan (\pi \lambda e^x ) )^{-1} )\sigma_2 - (\zeta_1 \zeta_3 - \zeta_2 (\tan (\pi \lambda e^x) )^{-1} )\sigma_3 )\\
 F_{14} = \frac{i}{2} (\sin (\pi \lambda e^x )^2 (-\pi \lambda e^x) ((-\zeta_2 (\tan (\pi \lambda e^x ))^{-1} + \zeta_1 \zeta_3 )\sigma_3 - (\zeta_1 \zeta_2 + \zeta_3 (\tan (\pi \lambda e^x ))^{-1} )\sigma_2 - (\zeta_2 \zeta_2 + \zeta_3 \zeta_3 )\sigma_1 )\\
 F_{24} = \frac{i}{2}(\sin (\pi \lambda e^x )^2 (-\pi \lambda e^x) ((\zeta_1 (\tan (\pi \lambda e^x ))^{-1}+ \zeta_2 \zeta_3 )\sigma_3 - (\zeta_1 \zeta_2 + \zeta_3 (\tan (\pi \lambda e^x ))^{-1} )\sigma_1 + (\zeta_1 \zeta_1 + \zeta_3 \zeta_3 )\sigma_2 )\\
 F_{34} = \frac{i}{2}(\sin (\pi \lambda e^x )^2 (-\pi \lambda e^x) ((\zeta_1 (\tan (\pi \lambda e^x ))^{-1} - \zeta_2 \zeta_3 )\sigma_2 + (\zeta_1 \zeta_3 + \zeta_2 (\tan (\pi \lambda e^x ))^{-1} )\sigma_1 + (\zeta_1 \zeta_1 + \zeta_2 \zeta_2 )\sigma_3 )
 \end{eqnarray}
 These solutions have now to be stereographically projected down to $R^4$.
 This is done by applying the usual formulas for stereographic projection to the pullbacked potentials $A_{\mu}^a$, where a is the Isospin index refering to the gauge group SU(2): $x_i = \frac{2u_i}{1 + u^2 } , i = 1,2,3$ The $x_i$, i = 1,2,3,4 are the four embeding coordinates of the $S^3$, the bundle space of the Hopf bundle. Due to the constraint $\Sigma x^2_i = 1$ ,
 $x_4$ is determined by the first three $x_i$. Hence the fourth independent coordinate on $S^4$ is given by the suspension parameter x: $x = \frac{2 u_4}{1 + u^2 }$. The constraint variable $x_4$ is given by $x_4 = \frac{1 - u^2}{1 + u^2}$ and the $u_i$, i = 1,2,3,4 are the coordinates of the $R^4$, on which the $S^4$ is projected. As an example we will calculate $A_4^1$.
 \begin{equation} A_4^1 = \frac{i}{\cosh (x)} \zeta_1 = \frac{2i}{\cosh (x)} (x_1 x_3 + x_2 x_4 )
 \end{equation}
 Plugging in the expressions for the stereographic projection given above yields the formula for the potential on $R^^4$.

 The new solution could also play a role in the Electro-Weak theory (Glashow-Salam-Weinberg-Theory), since its gauge group is $G_{GSW} = SU(2) \times U(1)$. (From now on we will write GSW for Glashow-Salam-Weinberg) Hence we have
 \begin{equation}
 \pi_4 (G_{GSW}) = \pi_4 (SU(2) \times U(1))
 \end{equation}
 \begin{equation}
  = \pi_4 (SU(2))
  \end{equation}
   and so
   \begin{equation}
   \fbox{$\pi_4 (G_{GSW}) = Z_2 $}
   \end{equation}
    Furthermore, because of the identity
    \begin{equation}
    \pi_n (S^2) = \pi_n (S^3)
    \end{equation}
    for n $\geq$ 3 ~\cite{Wu} But
    \begin{equation}
    S^2 = SU(2)/U(1)
    \end{equation}
     which is isomorphic to the vacuum configurations of the isotriplet Higgs scalars of the GSW theory.
\subsection{Spin structure and $\pi_4 (SU(2)) $ }
The difference of the two representant mappings of the two equivalence classes of $\pi_4 (SU(2)) $ can be described by the spin structures on the circles of  $S^4 $ which are the inverse images of our representant mappings above ~\cite{Hit}. Spin structures are classified by the first Stiefel Whitney class $H^1 (M, Z_2 )$ where M is the manifold under consideration. In our case these are the circles $S^1 $, which are the inverse images back from $SU(2) = S^3 $, covering $S^4 $. Similar to the Hopf bundle case, where the Hopf invariant counts the linking number of the circles covering $S^3 $. There is an analog in our $^4 $ case to be described below by cobordism theory. Now
\begin{equation}
H^1 (S^1 , Z_2 ) = Hom (H_1 (S^1 , Z_2 ))
\end{equation}
\begin{equation}
= Hom (\pi_1 (S^1 ) , Z_2 )
 \end{equation}
  The second isomorphism comes about by the Hurwitz isomorphism ~\cite{St} But this sequence links the classification of the spin structures on the collection of circles on $S^4 $ to the inverse images of the generic mappings $\phi :S^4 \rightarrow S^3 $ which make up the collection of circles covering $S^4 $. But these are the representants of the equivalence classes of
  \begin{equation}
  \fbox{$\pi_4 (SU(2)) = \pi_4 (S^3 )$}
  \end{equation}
  Hence the mappings $\phi : S^4 \rightarrow S^3 $ are classified by the spin structures on the generic inverse images, the circles covering $S^4 $ .
\subsection{Braids, Knots and $\pi_4 (SU(2))$}
According to the work of the South-Korean mathematician Jie Wu ~\cite{Wu} there is a link between a combinatoric group G(n) to be defined in the appendix and homotopy groups $\pi_n (S^2)$ for $n > 2$ and hence by the Hopf mapping a link with $\pi_n (S^3)$ and G(n) for $n > 2$. What he shows is, that Z(G(n)), the center of G(n), is isomorphic to the n-th homotopy group of the 2-sphere and so Z(G(n)) is isomorphic to the n-th homotopy group of the 3-sphere for $n > 2$ by the Hopf fibration. From our knowledge about especially $\pi_4 (S^3)$ we know now, that Z(G(n)) is is isomorphic to $Z_2$. The connection with Artins braid and pure braid group ~\cite{Ar} comes into the game by the further results, Wu found:
\\
Let $B_n$ be Artins braid group for n strands and $P_n$ the pure Artin braid group (see appendix for definitions). Then:
\begin{enumerate}
\item The set of fixed points of $B_n$ action on $G_n$, as a group, is isomorphic to the subgroup of $\pi_n (S^2 )$ consisting of elements of order two.
\item The set of fixed points of $P_n$ on $G(n)$, as a group, is isomorphic to $\pi_n (S^2 )$.
\end{enumerate}
They should also play a prominent role in the two dimensional non-linear sigma model, since the instanton solutions are directly linked with the the Hopf bundle (see above at the end of the section about the sigma model) Braid groups and knots played already a longer time in Yang-Mills theory, Quantum gravity and generally in quantum field theory a role (see ~\cite{Ka} ), but it seems that the new classical solution characterized by $\pi_4 (SU(2))$ gives a more natural link with this subject, while before it seemed to be introduced artificially. Closer examination of this link shall be the subject of further investigation later on.
\\
Another aspect, how knot theory might enter is the augmentation of the standard Yang-Mills action
\begin{equation}
S =\frac{1}{4} \int d^4 x FF
\end{equation}
 by a term
 \begin{equation}
 \frac{\theta}{32 (\pi)^2 }\int d^4 x F^* F
 \end{equation}
  where F is the Yang-Mills curvature and the augmentation term is the wellknown expression for the second Chern class times the factor $\frac{\theta}{4}$, which can be written as a total differential
\begin{equation}
\int F^* F = \int d^4 x ( \partial (A \wedge dA + A \wedge A \wedge A )
 \end{equation}
 But this is nothing but equal to
 \begin{equation}
 \int_{S^3} d\sigma (A \wedge dA + A \wedge A \wedge A )
 \end{equation}
 which is the Chern-Simon Lagrangian ~\cite{Wit} of which Witten ~\cite{Wi} showed that it gives a derivation of the Jones Polynomials ~\cite{Jo}, an invariant,  characterizing knots. Mind the first term in the Chern Simons Lagrangian (a so called secondary characteristic class ~\cite{Ch} )
 \begin{equation}
 l = \int_{S^3} d\sigma (A \wedge dA )
 \end{equation}
  which is nothing but the Hopf linking number for the Hopf fiber bundle giving the linking number of the inverse images of the Hopf mapping $\pi : S^3 \rightarrow S^2 $ . the inverse images are the fibers of the Hopf bundle, being circles, here on $S^3$, the bundle space of the Hopf bundle. So l gives us the linking number of the inverse images of the Hopf mapping. That means taking a U(1) gauge theory and hence the gauge potential A, being an element of the Lie algebra of the gauge group, abelian, the second term in the Chern Simons Lagrangian vanishes and we are left the expression for the Hopf linking number as the Lagrangian. (plus the conventional term $\int FF d^4 x$ ).

\subsection{Skyrme model and nonlinear $\sigma$ model }
Houghtom et al. ~\cite{Ho} made the suggestion to take the similarities of magnetic monopoles and Skyrmions literally and adapt an idea of Donaldson ~\cite{Do} to classify and determine t'Hooft magnetic monopole solutions by introducing rational maps $R: S^2 \rightarrow S^2 $. In a seminal paper Arafune et al. ~\cite{Ar} showed, that in a certain gauge, the tHooft-Polyakov monopoles ~\cite{tH} are topologically classified by the Brouwer degree, the mapping being the Higgs field
\begin{equation}
n = \frac{1}{8\pi e} \int_{S^3} \epsilon_{ijk} \epsilon_{abc} \phi^a \partial_j \phi^b \partial_k \phi^c (d^2 \sigma )_i
\end{equation}
where the $\phi^a$ are a triple of unit length Higgs fields because we look at a SO(3) Yang-Mills Higgs theory. The field theory is defined on $R^3$ but the origin can be deleted to avoid singularities. But $R^3$-0  is a deformation retract of $S^2 $. On the other side, the Higgs-field fullfills the condition $\phi^a \phi^a =$ const which defines a $S^2$ in field space. Hence the Higgs field defines a mapping
\begin{equation}
\phi^a : S^2 \rightarrow S^2
\end{equation}
 The Higgs field transforms under SO(3) as a vector and is left invariant by the subgroup SO(2). Hence all points reached by $\phi^a$ are equivalent to $\frac{SO(3)}{SO(2)} =S^2$. All mappings $\phi^a$ are classified by $\pi_2 (SO(3)/SO(2)) $ All this carries over in a one to one fashion to the two dimensional sigma model (or Heisenberg Ferromagnet) ~\cite{H}. All classical solutions with nontrivial Brouwer degree are given by rational functions, i.e., algebraic mappings
 \begin{equation}
 \phi^a = \frac{P(z)}{Q(z)}
 \end{equation}
  where z is the coordinate parametrizing the Riemann sphere = $S^2$. Hence, as in the case of monopoles due to the description of Donaldson, here, we have rational mappings. The Skyrmions are produced by suspending the mappings $\phi^a :S^2 \rightarrow S^2 $ to mappings U:$S^3 \rightarrow S^3$, U being the Skyrme function. ~\cite{Sk}. The author feels, that this is a more natural junction than with magnetic monopoles, also since the Skyrme model is in fact a nonlinear sigma model. So, Skyrmions should be gotten by suspension of spin wave solutions ~\cite{H}. Also, according to the article of Arafune et al., depending on the gauge, the magnetic charge of the t'Hooft monopole is carried either by the Higgs or by the gauge field. This makes t'Hooft monopoles to the opinion of the author interesting objects in a mathematical sense but it seems to him that they are unrealistic.
Since the new classical solution ~\cite{Ha} is generated by a mapping $\phi :S^4 \rightarrow S^3 $, which is in turn generated by the Hopf mapping of Hopf index 1 ~\cite{Ho}, it is natural to ask: What is the connection between the classical solution of Yang-Mills theory and Hopfions ~\cite{Fa}. Hopfions are solitons generated by Hopf mappings, gaining their stability from $\pi_3 (S^2) = Z$, that means, Hopf mappings are homotopically nontrivial. In contrast to, for instance, instantons, their topological charge is not the mapping degree, i.e., how often is $S^{2n-1}$ mapped on $S^n$, but by a topological invariant, called linking number, defined in the context of Hopf bundles by Hopf ~\cite{Ho}.

\section{The relationship of the new solution to Hopfions}
Hopfions are classical solutions of nonlinear field equations. The field of this nonlinear field theory is commonly named n and described by
\begin{equation}
n = (\sin (\theta ) \cos (\varphi ), \sin (\theta ) \sin (\varphi ), \cos (\theta)
\end{equation}
Thus n is subject to the cnstraint
\begin{equation}
n^2 = 1
\end{equation}
The angles $\theta$ and $\varphi$ are functions of $x_i$, i = 1,2,3, the coordinates of a $R^3$. Hence n is a mapping from $R^3$ to $S^2$. As in the case of instantons or the two dimensional non-linear sigma model ~\cite{H}, we demand that n has the same limit for $r \rightarrow \infty$ where r is the radius vector in $R^3$. This makes by one point compactification out of $R^3$ a three dimensional sphere $S^3$. Hence n is a mapping
\begin{equation}
n: S^3 \rightarrow S^2
\end{equation}
but this defines a Hopf bundle \cite{Ho} and this explains the name Hopfion for classical solutions of the non linear field theory of n. Hopfions are classical solutions owning a topological charge. But unlike instantons in Yang-Mills theory or the non-linear sigma model, their topological charge is not the mapping degree of n, but a topological invariant, called linking number. There is no mapping degree as in Yang-Mills or sigma model theory, since the dimensions of the manifolds are unequal ~\cite{Mil} The linking can be explained according to Hopf ~\cite{Ho} as follows: The inverse mapping of n, $n^{-1}$, maps a point p on $S^2$ into the fiber $S^1$ of the bundle space $S^3$. These fibers describe circles on the bundle space $S^3$ and the linking number counts the number, how often these circles are linked with each other. A differential topological expression of this number has been given by Whitehead ~\cite{Wh}
\begin{equation}
l = \int_{S^3} d\omega \wedge \omega
\end{equation}
where $\omega$ is the connection one form of the Hopf bundle. The stereographic projection down to $R^3$ is a geometric figure called ''Villarceau circle'' ~\cite{Bo}. The three components of the vector field n are the three coordinates $\zeta_i$, i = 1,2,3, of the $R^3$, in which the base space $S^2$ of our Hopf bundle is embedded. Identify these $\zeta_i$ with those used for the mapping $g : S^4 \rightarrow S^3$,
\begin{equation}
g(x,z ) = \exp[i\sigma^a \zeta^a \varphi (x) ]
\end{equation}
 where the $\zeta^a$ constitute the Hopf map
 \begin{equation}
 f:S^3 \rightarrow S^2
 \end{equation}
  with Hopf invariant 1 (Heinz Hopf, Collected papers, Springer, Berlin)
 used to construct the new classical solution in SU(2) Yang-Mills gauge theory ~\cite{Ha} and we have their connection to Hopfions. In the equations 11,12,and 13 of ~\cite{Ha} we gave an expression of the $\zeta_i$ in terms of four coordinates $x_i$, i = 1,2,3,4, which are the coordinates of $R^4$ being the space in which the bundle space $S^3$ of our Hopf bundle is embedded.
\begin{equation}
\zeta_1 = 2(x_1 x_3 + x_2 x_4 ) = \sin \theta \cos \varphi
\end{equation}
 \begin{equation}
 \zeta_2 = 2(x_1 x_4- x_2 x_3 ) = \sin \theta \sin \varphi
 \end{equation}
 \begin{equation}
  \zeta_3 = x_1^2 + x_2^2 - x_3^2 - x_4^2 = \cos \varphi
  \end{equation}
  The $x_i$ , i = 1,2,3,4, are the coordinates of the Euclidean space, in which $S^3$ is embedded. Hence we can replace the $\zeta_i$ in the mapping $g(x, x_i)$ by the Hopfion solution. since it is equally well a Hopf mapping. This opens up the possibility to create a machinery to produce different kinds of the new classical solution of SU(2) Yang-Mills theory, since any Hopf map with nontrivial linking number suspended will give a nontrivial representant of $\pi_4 (S^3)$, classifying the new solution ~\cite{Ha}.

\end{document}